\documentclass[12pt,a4paper]{article}
\usepackage{amsmath}
\usepackage{amsthm,amssymb}
\usepackage{subfigure}
\usepackage{graphicx} 
\usepackage{comment}

\newcommand{\beq}{\begin{equation}}
\newcommand{\eeq}{\end{equation}}

\newcommand{\allowedpairwalks}{\Omega}

\newcommand{\bottomcontactsites}{m_b (\varphi)}
\newcommand{\topcontactsites}{m_a (\varphi)}
\newcommand{\contactsites}{m (\varphi)}

\numberwithin{equation}{section}

\begin{document}
\title{Exact solution of asymmetric gelation between three walks on the square lattice}
\author{Aleksander L Owczarek$^1$ and Andrew Rechnitzer$^2$ \\[1ex]
	\footnotesize
	\begin{minipage}{9cm}
		Department of Mathematics,\\
		University of British Columbia,\\
		Vancouver V6T 1Z2, British Columbia, Canada.\\
		$^1$
		\texttt{aleks.owczarek@gmail.com}\\[1ex]
		$^2$
		\texttt{andrewr@math.ubc.ca}
	\end{minipage}
}

\maketitle

\begin{abstract}
	We find and analyse the exact solution of  a model of three different
	polymers with asymmetric contact interactions in two dimensions, modelling a
	scenario where there are different types of polymers involved. In particular, we find
	the generating function of  three directed osculating walks in star
	configurations on the square lattice with two interaction Boltzmann weights,
	so that there is one type of contact interaction between the  top pair of walks and a
	different interaction between the bottom pair of walks. These osculating
	stars are found to be the most amenable to exact solution using functional
	equation techniques in comparison to the symmetric case where three friendly
	walks in watermelon configurations were successfully solved with the same
	techniques.  We elucidate the phase diagram, which has four phases, and find
	the order of all the phase transitions between them. We also calculate the
	entropic exponents in each phase.

\end{abstract}
\newpage

\section{Introduction}

Models of polymer gelation have received interest over an extended period of time based upon understanding the phase behaviour of
systems of multiple polymers \cite{Elias2005}. Models of a small fixed number of polymers have been studied in their own right for various
reasons. For example, models of the unzipping
of DNA which naturally lead to the study of two polymers with interpolymer interactions have been a research focus \cite{essevaz-roulet1997a-a, lubensky2000a-a,
	lubensky2002a-a, orlandini2001a-a, marenduzzo2002a-a, marenduzzo2003a-a,
	marenduzzo2009a-a, owczarek2012exact, tabbara2014exact}.
Two varieties of models have the two polymers  modelled via either self-avoiding or directed
walk systems on lattices in two and three dimensions with various types of
contact interactions. The exact solution of
directed so-called friendly walkers, which can share edges and sites, on the square lattice with such interactions
\cite{owczarek2012exact, tabbara2014exact} has led to the extension of a key
combinatorial technique for lattice paths, the \emph{obstinate kernel method}
\cite{bousquet2002counting}.  More generally the binding of multiple polymers have been studied for some time \cite{Tabbara:2016dh,Jensen:2016kf,Beaton2021aa,Brak:2013hl,Essam2003ew,owczarek2017wu,bousquet2006three,guttmann2002lattice}.

The modelling of three polymers has considered in the context of polymer binding/zipping
\cite{Tabbara:2016dh}. One source of motivation was the search for models of Efimov-like states in  triple stranded DNA \cite{Maji:2010jq,Mura:2016dx}.
In \cite{Tabbara:2016dh} an exact solution was found for  three interacting friendly directed walks on the square lattice in the bulk.  Two distinct interaction parameters were introduced: one that acts on pairs of walks and one that acts when all three walks comes together.  The exact solution was found by the analysis of functional equations for the model's corresponding generating function by means of the obstinate kernel method. Without the triple walk interaction the model exhibits two phases which can be classified as free and gelated (or zipped), with the system exhibiting a second-order phase transition between these phases. The interactions between the different pairs of polymers were identical, modelling the situation where the homopolymers are all of the same type. It is of general interest then to consider a generalisation where the polymers may be different from one another so that the pairwise interaction may differ. It is such a model we consider in the work.

\begin{figure}[h!]
	\centering
	\includegraphics[width=250px]{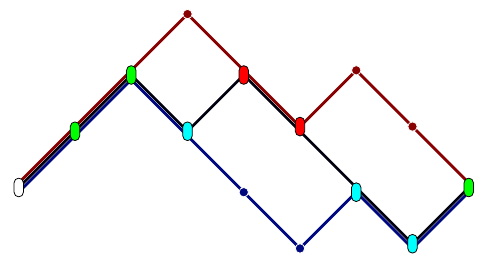}
	\caption[Three friendly watermelon]{Three friendly walks in a watermelon configuration as considered  previously \cite{Tabbara:2016dh}. }
	\label{3fw_figure_typical_config}
\end{figure}
In \cite{Tabbara:2016dh} the directed walk system consisted of three friendly walks (see Figure~\ref{3fw_figure_typical_config}) in which the walks may share sites and edges though are considered never to cross. They were specifically considered to start and end together. In our preliminary attempts to analyse a problem where the interactions between different pairs of walks were different we were unable to solve the corresponding functional equations as the breaking of the symmetry meant  it appeared we lacked sufficient equations to solve the model. Moreover, the D-finite nature of the generating function in the symmetric case meant we were unable to guess a solution in the more general case.

\begin{figure}[h!]
	\centering
	\includegraphics[width=250px]{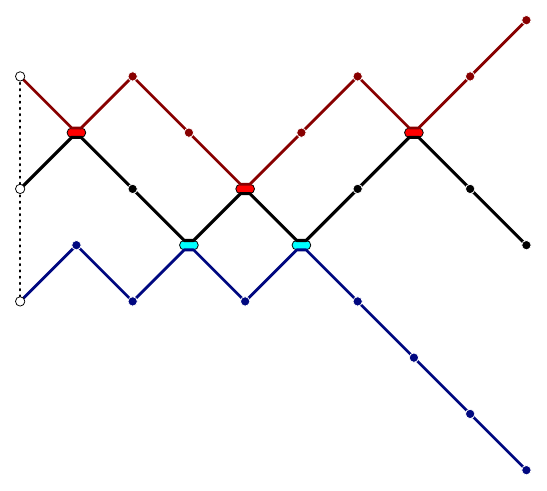}
	\caption[Three osculatng stars]{Three osculating walks in a star configuration as considered in this work. An example of an allowed configuration of length $n=9$. The walks start one lattice spacing apart (distance two) at coordinates $(0,0)$, $(0,2)$ and $(0,4)$, and end at $x=n$ with any allowed $y$. In this configuration the upper two walks end (rightmost)  two lattice spacing (distance four) apart whilst the bottom two walks end three lattice spacings (distance six) apart. Here, we have $m_a=3$  shared sites between the upper two walks and $m_b=2$ shared sites between the lower two walks. Thus, the overall Boltzmann weight for this configuration is $a^{3} b^{2}$.}
	\label{3ow_figure_typical_config}
\end{figure}
We turned to the study of other varieties of configuration in the hope that there may be a more amenable type whilst still allowing the  required  interactions. We noticed that both two and three osculating walks, where walks may share vertices but not edges, in star configurations (see Figure~\ref{3ow_figure_typical_config}) and the final endpoints are summed over gave algebraic solutions for the generating functions. As such it pointed to the idea that considering the asymmetric version of a similar three walk system would be a prudent line of attack. It turned out to be so and although we did not use the kernel method to solve the problem it was feasible to find the algebraic equation satisfied by the solution. Along the way we could elucidate the solution, provide the exact phase diagram and calculate all important exponents.

\section{Three Osculating Stars}
\label{3ow_model}
%
\subsection{Model definition}
In the Euclidean plane with coordinates $x$ and $y$ consider three directed walks on  the square integer lattice which has been rotated by $45^\circ$. The lattice has sites at coordinates $(x,y)=(2i,2j)$ and $(2i+1,2j+1)$ for $i,j\in \mathbb{Z}$. A square of the lattice is defined by the quartet of coordinates $(2i,2j)$, $(2i+1,2j+1)$, $(2i+1,2j-1)$ and $(2i+2,2j)$. The three directed walks consisting of an equal number of steps $n$ so that the total number of steps in a configuration is $3n$. All walks begin at $x=0$ with $y=0,2,4$, and end at $x=n$ with any heights. Walks do not cross. Moreover, walks only can take steps in either the north-east $(1,1)$ or south-east $(1,-1)$ direction along edges of the squares. Finally, any pair of  walks may share common site, however none of the walks are able to share common edges. Such walks are typically referred to as  \emph{osculating} walks. This contrasts  to the model considered in \cite{Tabbara:2016dh} where (infinitely) \emph{friendly} walks were considered where walks may share edges and sites without restriction, but may not cross. The other difference between the models is that in  \cite{Tabbara:2016dh}  the walks always ended at the same site (or otherwise close by at a fixed distance), known as a watermelon configuration; here we consider \emph{stars} where we sum over all possible end points. Let $\allowedpairwalks_3$ denote the class of allowed triple walks of \emph{any} length. An example of an allowable configuration is given in Figure~\ref{3ow_figure_typical_config}.
%
For any configuration $\varphi \in \allowedpairwalks_3$, we assign a
weight $a$ to the $m_a (\varphi)$ \emph{shared contact sites} and a weight $b$ to the $m_b (\varphi)$ \emph{shared contact sites} between the
top-to-middle and the middle-to-bottom walks respectively. Note, all three
walks cannot share the same site.  The partition function for our model consisting of $n$
triple steps is
%
\begin{equation}
	\label{3ow_partition_fn}
	Z_n(a,b) =\sum_{\varphi \in \allowedpairwalks_3, \\ |\varphi| = n}
	a^{\topcontactsites} b^{\bottomcontactsites},
\end{equation}
where $|\varphi|$ denotes the length of the configuration $\varphi$. The reduced free energy $\kappa(a,b)$ is given by
\begin{equation}
	\label{3ow_limiting free_energy}
	\kappa(a,b) = - \lim_{n \rightarrow \infty}\frac{1}{n} \log{Z_n(a,b)}.
\end{equation}
and generating function $G_3(a,b;z)$ for three walks is defined as
\begin{equation}
	\label{3ow_gen_fn}
	G_3(a,b;z) = \sum_{n=0}^{\infty} Z_n(a,b) z^n
\end{equation}
in the usual manner, where $z$ is conjugate to the length of the configuration.
Importantly, the relation between the free energy $\kappa$ and the radius of convergence of $G(a,b;z)$ is given by
\begin{equation}
	\label{tow_free_energy_rel}
	\kappa(a,b) = \log z_s (a,b),
\end{equation}
where $z_s (a,b)$ is the real and positive singularity of the generating function that is closest to the origin.  Moreover, it is expected that for any fixed $a$ and $b$  the partition function scales as
\begin{equation}
	\label{partfun-scale}
	Z_n (a,b) \sim A(a,b)\;    \mu(a,b)^n \; n^{\gamma -1},
\end{equation}
which defines the standard polymer exponent known as the \textit{entropic} exponent $\gamma$, the value of which is
expected to depend upon the phase or phase boundary at which the partition
function is evaluated \cite{gennes1979a-a,madras1993a-a,van2000statistical}. The \textit{growth rate} $\mu$ is a function of $a$
and $b$ and is given by
\begin{equation}
	\label{tow_free_energy_connnective}
	\mu (a,b) =  z_s (a,b)^{-1} = e^{-\kappa(a,b)}.
\end{equation}
From the generating function solution we will find the location of the singularity $z_s (a,b)$ which then effectively gives us the growth rate $\mu$ and free energy $\kappa$.

We show that there are four possible natural phases in our system which we will fully delineate in our analysis below. The phases are characterised by canonical order parameters related to the number of contacts between the walks and conjugate to the Boltzmann weights $a$ and $b$.
So, as such, we introduce the order parameters $\mathcal{A}(a,b)$ and $\mathcal{B}(a,b)$ being the limiting average density of contacts  between the top two and bottom two walks respectively as
\begin{equation}
	\mathcal{A}(a,b) = \lim_{n \rightarrow \infty} \frac{\langle m_a \rangle}{n} = a \frac{\partial \kappa}{\partial a},
\end{equation}
and
\begin{equation}
	\mathcal{B}(a,b) = \lim_{n\rightarrow \infty} \frac{\langle m_b \rangle}{n} = b \frac{\partial \kappa}{\partial b}.
\end{equation}
We can then characterise the possible phases in the following way. We say that the system is in a \emph{free} phase when
\begin{equation}
	\mathcal{A} =\mathcal{B} =0.
\end{equation}
A \emph{top zipped} phase with the top two walks only  bound together is indicated by the situation when
\begin{equation}
	\mathcal{A} > 0  \mbox{ with }  \mathcal{B} =0 ,
\end{equation}
whilst a \emph{bottom zipped} phase with the bottom two walks only  bound together is  indicated by the situation when
\begin{equation}
	\mathcal{B} > 0  \mbox{ with } \mathcal{A} =0.
\end{equation}
When both
\begin{equation}
	\mathcal{B} > 0 \mbox{ and } \mathcal{A} >0
\end{equation}
all three walks are bound, which we refer to as \emph{fully zipped}.

Phase transitions  are defined by  non-analytic behaviour of the free energy and so are  indicated by a non-analytic change in the singularity of the generating function. It is usual to define the standard polymer exponent $\alpha$ \cite{brak1993a-:a,van2000statistical} as related to the non-analyticity in the free energy $\kappa^{non}$ as
\begin{equation}
	\kappa^{non} \sim K\;  t^{2-\alpha}   \mbox{ as } t \rightarrow 0 ,
\end{equation}
where $t$ measures in Boltzmann weights $a$ and $b$  (or temperature) the distance to the phase transition and $K$ is a constant. This implies that the associated order parameter $\mathcal{M} = \mathcal{A,B}$ behaves as
\begin{equation}
	\mathcal{M} \sim M\;  t^{1-\alpha} \mbox{ as } t \rightarrow 0^+ .
\end{equation}
A standard scaling argument \cite{brak1993a-:a,van2000statistical} connects this to the scaling of the number of contacts evaluated exactly at the transition
\begin{equation}
	\langle m_{a,b} \rangle{(n)} \sim C \; n^\phi ,
\end{equation}
where
\begin{equation}
	\phi =  2-\alpha.
\end{equation}
Here which contact number $m_a$ or $m_b$ is considered is related to the order parameter(s) that are changing from zero to a non-zero value as the phase transition point is traversed.

\section{Functional equations for the generating function}
\label{3fw_functional_eqns}
We can establish a functional equation for $G(a,b;z)$ by considering the effect of appending a triplet of steps to the end of any given configuration $\varphi \in \allowedpairwalks_3$. To begin, we define $\allowedpairwalks_3(i,j)$ to be the class of triple osculating walks that consists of configurations with final top to middle walk distance $i$ and middle to bottom distance $j$, that still obey the osculating constraints. The full combinatorial class $\allowedpairwalks_3$ is then the union
\begin{equation}
	\allowedpairwalks_3 \equiv \bigcup_{i \geq 0, j \geq 0} \allowedpairwalks_3(i,j).
\end{equation}
Equipped with our refined combinatorial classes we can introduce its corresponding generating function $G(a,b;z)$ that encodes information about the number of steps and shared contacts for each configuration $\varphi \in \allowedpairwalks_3$. However, determining whether appending a triple-step onto a given configuration $\varphi$ results in a new and \emph{allowable} configuration (i.e. $\varphi$ remains in $\allowedpairwalks_3$) further requires knowledge of the \emph{final} step distances between the three walks. Hence, solely for the purpose of establishing our functional equation for $G(a,b;z)$, we additionally introduce two catalytic variables $r$ and $s$ to construct the expanded generating function $F(r,s,a,b;z)$ where
\begin{equation}
	\label{3fw_expanded_gen_fn}
	F(r,s,a,b;z) \equiv F(r,s) = \sum_{\varphi \in \allowedpairwalks_3} z^{|\varphi|}  r^{h(\varphi)/2} s^{f(\varphi)/2}  a^{\topcontactsites} b^{\bottomcontactsites},
\end{equation}
and again $z$ is conjugate to the length $|\varphi|$ of a configuration $\varphi \in \allowedpairwalks_3 $, $r$ and $s$ are conjugate to \emph{half} the distance $h(\varphi)$ and $f(\varphi)$ between the final vertices of the top to middle and middle to bottom walks respectively. For each $\varphi \in \allowedpairwalks_3$, powers of $r$ and $s$ in $F(r,s)$ track the final step distances between the three walks. Due to the allowed step directions, both $h(\varphi)$ and $f(\varphi)$ must always be even, ensuring that $F(r,s)$ contains only integer powers of $r$ and $s$. Thus, we consider $F(r,s)$ as an element of $\mathbb{Z}[r,s,a,b][[z]]$: the ring of formal power series in $z$ with coefficients in $\mathbb{Z}[r,s,a,b]$.
\par
We aim to solve $F(1,1,a,b;z) \equiv G_3(a,b;z)$ by establishing a functional equation for $F(r,s)$. Specifically, we construct a suitable mapping from $\allowedpairwalks_3$ onto itself by considering the effect of appending an \emph{allowable} triple-step onto a configuration, translating this map into its action on the generating function.

We then follow the procedure suitably modified found in \cite{Tabbara:2016dh} to construct the functional equation:

\begin{equation}
	\label{3ow_main_fn_eqn}
	\begin{aligned}
		K(r,s) F(r,s) & = rs - \left( \frac{(1+r)(r+s+rs)z}{rs} + \frac{1-b}{b} \right)F(r, 0)    \\
		              & \quad - \left( \frac{(1+s)(r+s+rs)z}{rs} + \frac{1-a}{a} \right) F(0,s) ,
	\end{aligned}
\end{equation}
where the \emph{kernel}, $K(r,s)$, is
\begin{equation}
	\label{3ow_kernel}
	K(r,s) \equiv K(r,s;z) = 1-\frac{z(r+1) (s+1) (r+s)}{rs}.
\end{equation}

\section{Exact solution of osculating stars}

\subsection{Two interacting osculating star walks}
We start by reproducing  the solution for two osculating walks with contact
interaction \cite{Fisher1984um,katori2001, guttmann2002lattice}. Analogous to
our definitions for the generating function of three walks we have the extended
generating function
\begin{equation}
	\label{2fw_expanded_gen_fn}
	F(s,a;z) \equiv F(s) = \sum_{\varphi \in \allowedpairwalks_2} z^{|\varphi|}  s^{d(\varphi)/2}  a^{\contactsites},
\end{equation}
where $\contactsites$ counts the number of contacts between the two walks and
$d(\varphi)$ the distance between the final vertices of the two walks. We want
the generating function for star configurations given by $G_2(a,z)=F(1,a;z)$.
Here $\allowedpairwalks_2$ denote the class of allowed pairs of walks of
\emph{any} length.

The functional equation for $F(s)$ is
\begin{equation}
	\label{2ow_main_fn_eqn}
	\begin{aligned}
		\left(1 - z(s+2+1/s) \right)F(s) & = s + \left( \frac{a-1}{a} - z(2+1/s) \right) F(0),
	\end{aligned}
\end{equation}
where \(s\) is now conjugate to the distance between the endpoints of the two walks. Note that the walks
start 1 lattice spacing apart.

This equation can be solved using the now
fairly standard approach of the kernel-method. Solving the kernel gives two roots
\begin{equation}
	s = \sigma_{\pm}(z) = \frac{1 - 2z \pm \sqrt{1-4z}}{2z}
\end{equation}
of which only one, \(\sigma_-(z)\), is combinatorial (being analytic at zero). Substituting this into the functional equation then eliminates
the unknown \(F(s)\) leaving a single equation for \(F(0)\), which yields:
\begin{equation}
	F(0;a,z) = \frac{a \sigma_-(z)^2}{az + \sigma_-(z)(1-a(1-2z))}.
\end{equation}
Back substitution into equation~\eqref{2ow_main_fn_eqn} and setting \(s=1\) then gives
\begin{equation}
	G_2(a,z) = \frac{1 + (3z-1)a}{2z(1+a(2z-1)+z^2a^2) \sqrt{1-4z}} - \frac{1+a(3z-1)+2z^2a^2}{2z(1 + a(2z-1) + z^2a^2)}.
\end{equation}
This generating function has two physical singularities; a square-root singularity when \(z=1/4\) and a simple pole
when \(z = \frac{\sqrt{a}-1}{a}\). These meet at \(a=4\) and the singularities coalesce giving
\begin{equation}
	G_2(4,z) = \frac{16z}{(3+4z)(1-4z)} -\frac{3}{2z(3+4z)\sqrt{1-4z}}  -  \frac{3}{2z(3+4z)}
\end{equation}
and the behaviour is dominated by a simple pole at \(z=1/4\) with a confluent one-on-square-root singularity.
The asymptotic behaviour of $m(a)$ the average number of
contacts between the two walks as a function of $a$ is given explicitly by
\begin{equation}
	m(a) = \langle m \rangle = \begin{cases}
		\frac{a}{(4-a)} + O(n^{-1})                & a<4     \\
		\frac{1}{\sqrt{\pi}} \cdot \sqrt{n} + O(1) & a=4     \\
		\frac{a-\sqrt{a}-2}{2(a-1)}\cdot n + O(1)  & a > 4 .
	\end{cases}
\end{equation}
Hence the associated order parameter
\begin{equation}
	\mathcal{M}(a) = \lim_{n \rightarrow \infty} \frac{\langle m \rangle}{n} = a \frac{\partial \kappa}{\partial a},
\end{equation}
is given by
\begin{equation}
	\mathcal{M}(a) = \begin{cases}
		0                           & a \leq 4 \\
		\frac{a-\sqrt{a}-2}{2(a-1)} & a > 4 .
	\end{cases}
\end{equation}
This implies that for $a<4$ there is a free phase where the two walks do not share a macroscopic number of sites whilst they are zipped together for $a>4$ sharing a non-zero macroscopic density of sites.
Our generating function leads us to provide the table of entropic exponents in Table~\ref{2ow_coeff_growth_table}.
\begin{table}[h!] \caption{\label{2ow_coeff_growth_table} The growth rate and entropic exponent for two walk stars.}
	\begin{center}
		\begin{tabular}{| l | c | c | }
			\hline
			Phase region              & $\mu$                  & $ \gamma $ \\
			\hline
			Free                      & 4                      & $1/2$      \\
			Zipped                    & $\frac{a}{\sqrt{a}-1}$ & $1$        \\
			\hline
			Free to Zipped transition & 4                      & $1$        \\
			\hline
		\end{tabular}
	\end{center}
\end{table}

Finally, we note that the phase transition is a continuous one with $\alpha=0$  (the order parameter decays linearly on approaching the transition) and $\phi=1/2$. There is a jump in the specific heat on traversing the transition at $a=4$.

\subsection{Three walks: symmetric interactions $a=b$}
We can explicitly solve the functional equation (\ref{3ow_main_fn_eqn})  for three osculating walks when $a=b$ following the same method applied by Bousquet-M\'{e}lou \cite{bousquet2002counting,bousquet2006three}. This reproduces the results found by Essam \cite{Essam2003ew}. When $a=b$ the physical variable conjugate to $a$ counts the total number of shared sites between pairs of osculating walks in three walk stars.

The kernel $K(r,s)$ does not depend explicitly on $a$ and $b$ whilst the function equation is
\begin{equation}
	\label{3ow_sym_fn_eqn}
	\begin{aligned}
		K(r,s) F(r,s) & = r^2 - \left( \frac{(1+r)(r+s+rs)z}{rs} + \frac{1-a}{a} \right)F(r, 0)   \\
		              & \quad - \left( \frac{(1+s)(r+s+rs)z}{rs} + \frac{1-a}{a} \right) F(0,s) .
	\end{aligned}
\end{equation}
We immediately notice the symmetry \(r \leftrightarrow s\), which facilitates the solution and that this symmetry is broken when \(a \neq b\).

The generating function $G_3(a,a;z)=F(1,1,a,a,z)$ is
\begin{equation}
	\label{symmetricsoln}
	\begin{aligned}
		G_3(a,a;z) & =
		\frac{\left(6 a z -a +1\right) \left(a z +1\right) \left(a z -1\right) }{4 z^{2} \left(a^{2} z -a +2\right) \left(4 a^{2} z^{2}+4 a z -a +1\right)}
		\cdot \sqrt{1-4z}   \\
		           & +  \frac{
			-8 a^{4} z^{4}+2 a^{2} \left(a -20\right) z^{3}+a \left(a +7\right) \left(a -4\right) z^{2}+\left(10 a -4\right) z -a +1
		}{4 z^{2} \left(a^{2} z -a +2\right) \left(4 a^{2} z^{2}+4 a z -a +1\right)}.
	\end{aligned}
\end{equation}
We immediately notice that this is an algebraic function and this contrasts to the D-finite non-algebraic solution for the generating function for three friendly walks in a watermelon configuration found in \cite{Tabbara:2016dh}. This provides a pointer to why we could extend the current model to the asymmetric solution below whilst could not  similarly extend the work in \cite{Tabbara:2016dh}.

The analysis of this generating function is very similar to that of the 2-walk problem. When \(a\) is small, the asymptotics are dominated by the
square-root singularity at \(z = 1/8\). When \(a\) is large, the problem is dominated by a simple pole at \(z = \frac{a-2}{a^2}\). We note that there is
an additional pole at \(z = \frac{\sqrt{a}-1}{2a} \) but it is dominated by the other. All three singularities coalesce when \(a=4\) and the
generating function simplifies to
\begin{equation}
	\begin{aligned}
		G_3(4,4;z) & =
		\frac{1}{1-8z} +
		\frac{3 \left(1-4 z \right) \left(4 z +1\right) }{8z^2 \left(8 z +3\right) \sqrt{1-8 z}}
		-\frac{3 \left(8 z^{2}+4 z +1\right)}{8 z^2 \left(8 z +3\right)}
	\end{aligned}
	\label{eq: aeqb gf}
\end{equation}
and the system is dominated by a simple pole at \(z=1/8\). Hence the singularity closest to the origin is given by
\begin{equation}
	\label{eq: aeqb zc}
	z_c =\begin{cases}
		\frac{1}{8}     & a \leq 4 \\
		\frac{a-2}{a^2} & a > 4.
	\end{cases}
\end{equation}

The asymptotic behaviour of average total number of contacts $m(a)$ between pairs of osculating walks as a function of $a$ is given by
\begin{equation}
	\label{eq: density aeqb}
	m(a) = \begin{cases}
		\frac{a(192 - 8a - a^2)}{(4-a)(64-a^2)} + O(n^{-1}) & a<4     \\
		\frac{3}{\sqrt{\pi}} \cdot \sqrt{n} + O(1)          & a=4     \\
		\frac{a-4}{a-2} \cdot n + O(1)                      & a > 4 .
	\end{cases}
\end{equation}
Hence the associated order parameter is given by
\begin{equation}
	\mathcal{M}(a) = \begin{cases}
		0                           & a \leq 4 \\
		\frac{a-4}{a-2} & a > 4 .
	\end{cases}
\end{equation}

Once again the transition is continuous with  $\alpha=0$ and $\phi=1/2$. Also, as with two walk stars when $a<4$ there is a free phase where the none of the pairs of walks share a macroscopic number of sites whilst they are all zipped together (fully zipped) for $a>4$ sharing a non-zero macroscopic density of sites.
Our generating function leads us to provide the table of entropic exponents in Table~\ref{3ow_coeff_growth_table}.
\begin{table}[h!] \caption{\label{3ow_coeff_growth_table} The growth rates and entropic exponents for three walk stars with symmetric interactions.}
	\begin{center}
		\begin{tabular}{| l | c | c |}
			\hline
			Phase region          & $\mu$             & $ \gamma $ \\
			\hline
			Free                  & 8                 & $-1/2$     \\
			Fully Zipped          & $\frac{a^2}{a-2}$ & $1$        \\
			\hline
			Free to Fully  Zipped & 8                 & $1$        \\
			\hline
		\end{tabular}
	\end{center}
\end{table}

\subsection{Three walks: asymmetric interactions for stars $r=s=1$}

We have not been able to use the kernel method to directly solve the above equation~\eqref{3ow_main_fn_eqn} for the watermelon case, namely \(r=s=0\), nor for general \(r,s\) by using any variant of the kernel method.
We believe that this is due to the breaking of the \(r \leftrightarrow s\) symmetry when \(a\neq b\).
We were, however, able to determine a quartic equation satisfied by the generating function of the star case, namely \(r=s=1\) by analytically guessing the solution and then confirming that it is indeed the solution. Given the solution of the symmetric problem is algebraic, and, in particular quadratic, there is some hope that the solution of the asymmetric problem might satisfy a  low degree algebraic equation. We note that this is precisely the case for a related problem - Kreweras walks - in which the symmetric problem satisfies a polynomial of degree 6 \cite{beaton2019a-:a} and the asymmetric problem satisfies a polynomial of degree 12 \cite{beaton2019a-:a,beaton2019b-:a,beaton2021b-:a}. Consequently we searched for low degree algebraic equations.

We note that Bousquet-M\'{e}lou \cite{bousquet2006three} derived a quadratic
solution in the non-interacting case \(r=s=a=b=1\). We found that for \(r=s=1\)
and general \(a=b\) the generating function also satisfied a quadratic. For
particular values of \(a \neq b\) (for example \( (a,b) = (1,2),(2,3),(2,4),
(3,4) \) and so on ) we generated long series and were able to guess quartic
equations using the Ore Algebras package \cite{kauers2015} for the Sage computer
algebra system. By
computing these quartics at sufficient particular values of \(a \neq b\) we
were able to construct the general \(a\neq b\) equation.  We give the quartic
equation in the appendix for the generating function $G\equiv G_3(a,b;z)$: it
is of the standard form
\begin{equation}
	\label{3ow-asym-quartic}
	\sum_{j=0}^4 c_j(a,b;z) G^j = 0 ,
\end{equation}
where the coefficients $c_j(a,b;z)$ are polynomials in $a$, $b$ and $z$. Hence
we could write a closed form solution for $G_3(a,b;z)$ is terms of the
classical solution of a quartic. It would be unwieldy and we can provide the
asymptotic results required by directly analysing the quartic. Important in this regard is that the
coefficient of $G^4$ is
\begin{equation}
	\label{3ow-asym-g4coeff}
	c_4(a,b;z) = 4 z^6 (4 b^2 z^2 + 4 b z - b + 1)(4a^2z^2 + 4az - a + 1)(a^2b^2z^2 + 2abz - ab + a + b)^2 .
\end{equation}

\section{Phase diagram and exponents}
As described in the definition of the model, its natural quantities, the two
candidate order parameters, point to four possible phases depending on whether
those two order parameters are non-zero or not.

Given that non-analyticities in the free energy are governed by the change of
singularities closest to the origin of the generating function the location of
those possible singularities are central to delineating the possible phases.
Since we know the quartic satisfied by the generating function, we can, in principle,
compute the generating function in closed form. Unfortunately, that proves to be
rather unwieldy for general \(a,b\), but is useful for confirming the location of singularities
for particular fixed choices of \(a,b\). Instead, we locate
possible singularities by examining its discriminant and the zeros of the leading
coefficient. We can also locate potential singularities by converting the algebraic
equation to a (second order) linear differential equation and examining the zeros
of its highest order term. In so doing, we find four independent singularities
($z_{\text{free}}, z_{\text{top}}, z_{\text{bottom}}, z_{\text{fully}} $) of the generating function which
correspond to these four phases when they are the singularity closest to the
origin at some value of $a$ and $b$:
\begin{equation}
	\label{3ow-sings}
	z_c =\begin{cases}
		z_{\text{free}} =\frac{1}{8}             \\
		z_{\text{top}}= \frac{\sqrt{a}-1}{2a}    \\
		z_{\text{bottom}}= \frac{\sqrt{b}-1}{2b} \\
		z_{\text{fully}} = \frac{-1 + \sqrt{(a-1)(b-1)}}{ab} .
	\end{cases}
\end{equation}
Notice that when \(b=a\) we recover the results in equation~\eqref{eq: aeqb zc}. Additionally, note that
\(z_{\text{free}}, z_{\text{top}}, z_{\text{bottom}}\) are precisely half the critical point of the two-walk model; this is commensurate with adding
a third (effectively) non-interacting directed walk to the system.

From this we can find the phase boundaries by looking where the singularities above are equal.
\begin{itemize}
	\item The singularities $z_{\text{free}}$ and $z_{\text{top}}$ meet  when \(\frac{1}{8} = \frac{\sqrt{a}-1}{2a} \) which happens at \(a=4\) and \(b \leq 4\), which we shall label $t_{1}$;
	\item The singularities $z_{\text{free}}$ and $z_{\text{bottom}}$ meet when \(\frac{1}{8} = \frac{\sqrt{b}-1}{2b} \) which happens at \(b=4\) and \(a \leq 4\), which we shall label $t_{2}$;
	\item The singularities $z_{\text{top}}$ and $z_{\text{fully}}$ meet, which we shall label $t_{t}$, when
	      \begin{equation}
		      \frac{\sqrt{a}-1}{2a}
		      =  \frac{-1 + \sqrt{(a-1)(b-1)}}{ab}
	      \end{equation}
	      which happens along the curve
	      \begin{equation}
		      a = \frac{b^2}{(b-2)^2}  \quad \mbox{ for } b < 4 ;
	      \end{equation}
	\item The singularities $z_{\text{bottom}}$ and $z_{\text{fully}}$ meet, which we shall label $t_{b}$,  when
	      \begin{equation}
		      \frac{\sqrt{b}-1}{2b}
		      =  \frac{-1 + \sqrt{(a-1)(b-1)}}{ab}
	      \end{equation}
	      which happens along the curve
	      \begin{equation}
		      b = \frac{a^2}{(a-2)^2} \quad \mbox{ for } a < 4 .
	      \end{equation}
\end{itemize}
We note that for large $a$ the transition boundary $t_{t}$ is asymptotic to the line $b=2$ and so there is utility in inverting the curve $a = \frac{b^2}{(b-2)^2} $ as
\begin{equation}
	b_c(a) = \frac{2\sqrt{a}}{\sqrt{a}-1} \quad \mbox{ for } a > 4.
\end{equation}
Similarly, the phase transition boundary $t_{b}$ occurs when
\begin{equation}
	a_c(b) = \frac{2\sqrt{b}}{\sqrt{b}-1} \quad \mbox{ for } b > 4.
\end{equation}

\begin{figure}[h!]
	\centering
	\includegraphics[width=300px]{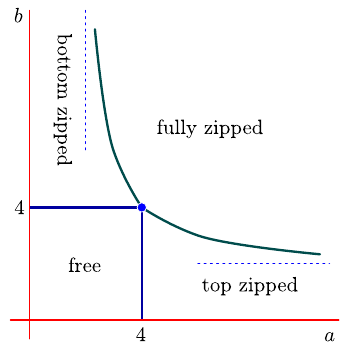}
	\caption[Phase Diagram]{The phase diagram of our model. The dashed lines indicate the asymptotes of the phase boundaries along \(b=2\) for large \(a\) and \(a=2\) for large \(b\).}
	\label{3ow_phase_diagram}
\end{figure}

Hence we deduce that the closest singularity to the origin is given by
\begin{align}
	\label{3ow_asym_loc_sing}
	z_c(a,b) & =\begin{cases}
		            z_{\text{free}} =\frac{1}{8}                              & \mbox{ when } a <4 \mbox{ and } b <4                                         \\
		            z_{\text{top}}(a) = \frac{\sqrt{a}-1}{2a}                 & \mbox{ when } a > 4  \mbox{ and }  0 \leq b<  \ \frac{2\sqrt{a}}{\sqrt{a}-1} \\
		            z_{\text{bottom}}(b) = \frac{\sqrt{b}-1}{2b}              & \mbox{ when } 0\leq a <    \frac{2\sqrt{b}}{\sqrt{b}-1}  \mbox{ and }  b> 4  \\
		            z_{\text{fully}}(a,b) = \frac{-1 + \sqrt{(a-1)(b-1)}}{ab} & \mbox{ otherwise. }
	            \end{cases}
\end{align}
These give us the growth constant $\mu$ and free energy $\kappa$ in each phase. Given these singularities we can now analysis the scaling of the generating function/ partition function and the two types of contact (hence the order parameters). Our phase diagram in Figure~\ref{3ow_phase_diagram}  is derived from the singularity structure given in equation~\eqref{3ow_asym_loc_sing} and our analysis below

\subsection{Free phase}
For low \(a,b\) when $a<4, b<4$ the dominant singularity is $z_c = z_{\text{free}}= \frac{1}{8}$. In this region we find that
\begin{equation}
	Z_n \sim 8^n \,n^{-3/2}.
\end{equation}
Leading to $\mu=8$ and $\gamma=-1/2$.

In this phase we have calculated that
\begin{equation}
	\begin{aligned}
		m_a (a,b) & = \frac{a(ab + 8b - 192)}{2(4-a)(ab - 64)} + O(n^{-1}), \\
		m_b (a,b) & = \frac{b(ab + 8a - 192)}{2(4-b)(ab - 64)} + O(n^{-1}).
	\end{aligned}
\end{equation}
so that
\begin{equation}
	\mathcal{A}=\mathcal{B}=0 \mbox{ when } a <4 \mbox{ and } b <4,
\end{equation}
which could be deduced by the constancy of the free energy in this region.
So we label this as a \textit{Free} phase where the three walks are effectively repulsive and share few (a finite number)  sites with one another.
Notice that when \(b=a\) we recover the \(a<4\) case in equation~\eqref{eq: density aeqb}.

\subsection{Partially zipped phase: Top Zipped}
The singularity in this phase arises from the  factor of \( (a^2z^2+2az-a+1) \)
in the coefficient $c_4$ of \(G^4\) in the quartic. In the region ($a > 4$  and   $0
	\leq  b <  \ \frac{2\sqrt{a}}{\sqrt{a}-1} $) where
$z_{\text{top}}=\frac{\sqrt{a}-1}{2a}$ dominates we have a phase where the free energy
is independent of $b$ so $\mathcal{B}=0$ but
\begin{equation}
	\mathcal{A} = \frac{a}{z_{\text{top}}(a) }\frac{\partial z_{\text{top}}(a) }{\partial a} = \frac{a-\sqrt{a}-2}{2(a-1)}.
\end{equation}
So in this phase we expect the top two walks share a macroscopic number of
sites osculating frequently effectively zipped together whilst the bottom walk
is repelled from the upper two walks. We refer to this phase as \textit{Top Zipped}.

In this region we find that
\begin{equation}
	Z_n \sim z_{\text{top}}^{-n} \; n^{-1/2}.
\end{equation}
Leading to
\begin{equation}
	\mu=\frac{1}{z_{top }} = \frac{2a}{\sqrt{a}-1}
\end{equation}
and $\gamma= 1/2$.

In this region we have
\begin{equation}
	\begin{aligned}
		m_a(a,b) & = \frac{a-\sqrt{a}-2}{2(a-1)} \cdot n + o(n), \\
		m_b(a,b) & = O(1).
	\end{aligned}
\end{equation}
We expect that \(m_b(a,b) = const + o(1)\) in this region, however we have not been
able to compute more precise value of this statistic. Although there is
no theoretical impediment to this calculation, we have found that various
computer algebra systems have been unable to expand \(G_3(a,b;z)\) about
\(z=z_{\text{top}}(a)\). Despite this we were able to guess the expressions but it is so convoluted it appears that they are too complicated to be
manipulated readily.

\subsection{Partially zipped phase: Bottom Zipped}
In the region ($0  \leq  a<  \ \frac{2\sqrt{b}}{\sqrt{b}-1} $ and $b > 4$ )
where $z_{bot}$ dominates we have a phase where the free energy is independent
of $a$ so $\mathcal{A}=0$ but
\begin{equation}
	\mathcal{B} = \frac{b}{z_{bot}(b) }\frac{\partial z_{bot}(b) }{\partial b} = \frac{b-\sqrt{b}-2}{2(b-1)}.
\end{equation}
So in this phase we expect the bottom two walks share a macroscopic number of
sites osculating frequently effectively zipped together whilst the top walk is
repelled from the lower two walks. We refer to this phase as \textit{Bottom Zipped}.

In this region we find that
\begin{equation}
	Z_n \sim z_{bot}^{-n} \; n^{-1/2}.
\end{equation}
Leading to
\begin{equation}
	\mu=\frac{1}{z_{bot}} =  \frac{2b}{\sqrt{b}-1}
\end{equation}
and $\gamma= 1/2$.

In this region we have
\begin{equation}
	\begin{aligned}
		m_a(a,b) & = O(1)    ,                                   \\
		m_b(a,b) & = \frac{b-\sqrt{b}-2}{2(b-1)} \cdot n + o(n). \\
	\end{aligned}
\end{equation}
Again, we expect that \(m_a(a,b) = const + o(1)\) in this region, but we have been unable to further elucidate the value  of this statistic simply.

\subsection{Fully Zipped}
Finally, when both \(a,b\) are large  there is a simple pole and it is given
by a zero of the leading coefficient in the algebraic equation; namely,
\begin{equation}
	0 = a^2b^2z^2 + 2abz - ab + a + b ,
\end{equation}
which gives
\begin{equation}
	z_c = z_{\text{fully}}=  \frac{-1 + \sqrt{(a-1)(b-1)}}{ab},
\end{equation}
and, moreover, the simple pole in the generating function results in the scaling of the partition function as
\begin{equation}
	Z_n \sim z_{\text{fully}}^n .
\end{equation}
Leading to
\begin{equation}
	\mu(a,b)=\frac{1}{z_{\text{fully}}(a,b)}
\end{equation}
and $\gamma=1$.

Finally, we calculate the order parameters, both of which are non-zero which implies that all three walks are zipped together as hence we refer to this phase a \textit{Fully Zipped}:
\begin{equation}
	\begin{aligned}
		\mathcal{A} (a,b) & =     \frac{a}{z_{\text{top}}(a) }\frac{\partial z_{\text{top}}(a) }{\partial a}
		                  & = \frac{1}{2(a-1)}\left(a-2 + \frac{a}{1+\sqrt{(a-1)(b-1)}} \right)        ,     \\
		\mathcal{B} (a,b) & =     \frac{b}{z_{bot}(b) }\frac{\partial z_{bot}(b) }{\partial b}
		                  & = \frac{1}{2(b-1)}\left(b-2 + \frac{b}{1+\sqrt{(a-1)(b-1)}} \right) .
	\end{aligned}
\end{equation}
Notice that \( \mathcal{A}(a,a)+\mathcal{B}(a,a) = \frac{a-4}{a-2}\) and so recovers the large \(a\) behaviour given
in equation~\eqref{eq: density aeqb}.

\subsection{Phase transition boundary scaling}
We have also calculated the scaling of the number of the two types of contacts on each of the phase boundaries.
\begin{itemize}
	\item For the phase boundary $t_{1}$, when $a=4$ and $b<4$ we have
	      \begin{equation}
		      \begin{aligned}
			      Z_n & \sim 8^n n^{-3/4}              ,                                      \\
			      m_a & \sim \frac{\pi\sqrt{2}}{4 \Gamma(3/4)^2}\cdot\sqrt{n} + O(n^{-1/2}) , \\
			      m_b & \sim \frac{b(40-b)}{2(4-b)(16-b)} + O(n^{-1}),
		      \end{aligned}
	      \end{equation}
	      which implies that $\mu = 8$, $\gamma = 1/4$ and $\phi=1/2$,
	\item For the phase boundary $t_{2}$, when $a<4$ and $b=4$ we have
	      \begin{equation}
		      \begin{aligned}
			      Z_n & \sim 8^n n^{-3/4}          ,                                         \\
			      m_a & \sim \frac{a(40-a)}{2(4-a)(16-a)} + O(n^{-1})   ,                    \\
			      m_b & \sim \frac{\pi\sqrt{2}}{4 \Gamma(3/4)^2}\cdot\sqrt{n} + O(n^{-1/2}),
		      \end{aligned}
	      \end{equation}
	      which implies that $\mu = 8$, $\gamma = 1/4$ and $\phi=1/2$,
	\item For the phase boundary $t_{t}$, when $b = \frac{2\sqrt{a}}{\sqrt{a}-1}$ and  $a > 4$ we have
	      \begin{equation}
		      \begin{aligned}
			      Z_n & \sim \left( \frac{2a}{\sqrt{a}-1} \right)^n n^0, \\
			      m_a & \sim \frac{a-\sqrt{a}-2}{2(a-1)} \cdot n + O(1), \\
			      m_b & \sim O(\sqrt{n}) + O(n^{-1/2}),
		      \end{aligned}
	      \end{equation}
	      which implies that $\mu = \frac{2a}{\sqrt{a}-1}$ and $\gamma =1 $. As before, we have not been able to compute more detailed
	      asymptotics for \(m_b\) along this boundary.
	\item For the phase boundary $t_{b}$, when $a = \frac{2\sqrt{b}}{\sqrt{b}-1}$ for  $b > 4$ we have
	      \begin{equation}
		      \begin{aligned}
			      Z_n & \sim \left( \frac{2b}{\sqrt{b}-1} \right)^n n^0, \\
			      m_a & \sim O(\sqrt{n}) + O(n^{-1/2})  ,                \\
			      m_a & \sim \frac{b-\sqrt{b}-2}{2(b-1)} + O(1),
		      \end{aligned}
	      \end{equation}
	      which implies that $\mu = \frac{2b}{\sqrt{b}-1}$ and $\gamma = $1. As before, we have not been able to compute more detailed
	      asymptotics for \(m_a\) along this boundary.

	\item For special critical point at $a=b=4$ we have
	      \begin{equation}
		      \begin{aligned}
			      Z_n & \sim 8^n n^0                       ,                      \\
			      m_a & \sim \frac{3}{2\sqrt{\pi}} \cdot \sqrt{n} + O(n^{-1/2}) , \\
			      m_b & \sim \frac{3}{2\sqrt{\pi}} \cdot \sqrt{n} + O(n^{-1/2}),
		      \end{aligned}
	      \end{equation}
	      which implies that $\mu =  8$ and $\gamma = 1$ with $\phi=1/2$. Notice that \(m_a+m_b\) gives the \(a=4\) result in \eqref{eq: density aeqb}.
\end{itemize}

\subsection{Results summary}
The phase diagram is delineated by equation~\eqref{3ow_asym_loc_sing}. We report that all of the phase transitions between these phases in our model are second-order with \(\alpha = 0\) and $\phi=1/2$.
We then summarize the growth rates/free energy and entropic exponents in Tables~\ref{3ow_asym_coeff_growth_table}:
\begin{table}[h!] \caption{\label{3ow_asym_coeff_growth_table} The growth rates (growth rates give the free energies) and entropic exponent for three walk stars with asymmetric interactions. In the top part of the table each of the primary phases are listed whilst in the bottom part of the table the phases boundaries are listed.}
	\begin{center}
		\begin{tabular}{| l | c | c| }
			\hline
			Phase regions                              & $\mu$                             & $\gamma $ \\
			\hline
			free                                      & 8                                 & $-1/2$    \\
			partially zipped (top)                    & $ \frac{2a} {\sqrt{a}-1}$         & $1/2$     \\
			partially zipped (bottom)                 & $ \frac{2b} {\sqrt{b}-1}$         & $1/2$     \\
			fully zipped                              & $\frac{ab}{\sqrt{(a-1)(b-1)}-1} $ & $1$       \\
			\hline
			Phase boundaries                              & $\mu$                             & $\gamma $ \\
			\hline
			free to partially zipped  (top)           & 8                                 & $1/4$     \\
			free to partially zipped   (bottom)       & 8                                 & $1/4$     \\
			free to fully zipped                      & 8                                 & $1$       \\
			partially zipped (top) to fully zipped    & $ \frac{2a} {\sqrt{a}-1}$         & $1$       \\
			partially zipped (bottom) to fully zipped & $ \frac{2b} {\sqrt{b}-1}$         & $1$       \\

			\hline
		\end{tabular}
	\end{center}
\end{table}

\pagebreak

\section{Conclusion}

We have ascertained the exact solution of  a model of three directed osculating walks with asymmetric contact interactions in star configurations. We expect that similar models such as friendly walks and watermelon configurations will behave similarly with  similar phase diagrams though the entropic exponents will differ for watermelon type configurations. We have delineated the order of the transitions and the associated exponents, especially the entropic exponents at all points in the phase diagram. We note that the values of $1/4$ on two of the phase boundaries are somewhat novel. At high temperatures the system is in a free state where the three polymers repel each other. At  low temperatures the system is in a bound state where all three polymers are bound together. Given any asymmetry there is a state at intermediate temperatures where two of the polymers are bound whilst the third polymer is repelled by the other pair. Regarding the method of exact solution of the model considered we note that we moved to the osculating case after not being able to solve the friendly model with asymmetric interaction though the symmetric model can be solved. A further understanding of the reasons for this and why the watermelon configuration is less amenable to solution would be most interesting.

\section*{Acknowledgements}
Financial support from the Australian Research Council through its Discovery Projects scheme is gratefully acknowledged.
Financial support from the Natural Sciences and Engineering Research Council of
Canada (NSERC) though its Discovery Program is gratefully acknowledged by the
authors. ALO thanks the Mathematics Departments at University of British Columbia, the University of Melbourne and Monash University respectively for hospitality.

\section{Appendix --- Equation for asymmetric stars}

The following is the quartic we found to be satisfied by the generating function $G\equiv G_3(a,b;z)$.
\begin{align*}
	0 =\;\; & G^4 \cdot 4 z^6 (4 b^2 z^2 + 4 b z - b + 1)(4a^2z^2 + 4az - a + 1)(a^2b^2z^2 + 2abz - ab + a + b)^2                                     \\
	        & + G^3 \cdot 8z^4(4b^2z^2 + 4bz - b + 1)(4a^2z^2 + 4az - a + 1) \cdot                                                                    \\
	        & \qquad (a^2b^2z^2 + 2abz - ab + a + b)(a^2b^2z^3 + 5abz^2 - abz + 2az + 2bz - 1)                                                        \\
	        & +G^2 \cdot  z^2 \Big(
	96 a^{6} b^{6} z^{10}+48 a^{5} b^{5} \left(2 a +2 b +23\right) z^{9}                                                                              \\
	        & -4 a^{4} b^{4} \left(6 a^{2} b +6 a \,b^{2}-6 a^{2}+41 a b -6 b^{2}-378 a -378 b -600\right) z^{8}                                      \\
	        & -4 a^{3} b^{3} \left(9 a^{2} b^{2}+120 a^{2} b +120 a \,b^{2}-172 a^{2}-99 a b -172 b^{2}-1080 a -1080 b \right) z^{7}                  \\
	        & +a^{2} b^{2} \left(5 a^{3} b^{3}+51 a^{3} b^{2}+51 a^{2} b^{3}-160 a^{3} b -679 a^{2} b^{2}-160 a \,b^{3}+104 a^{3}-1732 a^{2} b\right. \\
	        & \left. -1732 a \,b^{2}+104 b^{3}+2904 a^{2}+6040 a b +2904 b^{2}\right) z^{6 }                                                          \\
	        & +2 a b \left(71 a^{3} b^{3}+109 a^{3} b^{2}+109 a^{2} b^{3}-592 a^{3} b -1845 a^{2} b^{2}-592 a \,b^{3}+432 a^{3}+1402 a^{2} b\right.   \\
	        & \left. +1402 a \,b^{2}+432 b^{3}\right) z^{5}                                                                                           \\
	        & +\left(-11 a^{4} b^{4}-4 a^{4} b^{3}-4 a^{3} b^{4}+127 a^{4} b^{2}+987 a^{3} b^{3}+127 a^{2} b^{4}-208 a^{4} b -1579 a^{3} b^{2}\right. \\
	        & \left. -1579 a^{2} b^{3}-208 a \,b^{4}+96 a^{4}+432 a^{3} b +452 a^{2} b^{2}+432 a \,b^{3}+96 b^{4}\right) z^{4}                        \\
	        & -4 a b \left(32 a^{2} b^{2}-77 a^{2} b -77 a \,b^{2}+50 a^{2}-12 a b +50 b^{2}+52 a +52 b \right) z^{3}                                 \\
	        & +\left(7 a^{3} b^{3}-25 a^{3} b^{2}-25 a^{2} b^{3}+34 a^{3} b -97 a^{2} b^{2}+34 a \,b^{3}-16 a^{3}+194 a^{2} b\right.                  \\
	        & \left. +194 a \,b^{2}-16 b^{3}-48 a^{2}-112 a b -48 b^{2}\right) z^{2}                                                                  \\
	        & +\left(22 a^{2} b^{2}-46 a^{2} b -46 a \,b^{2}+24 a^{2}+30 a b +24 b^{2}-4 a -4 b \right) z                                             \\
	        & -\left(b -1\right) \left(a -1\right) \left(a b -a -b -4\right)	\Big)                                                                    \\
	        & + G^1 \cdot  (a^2b^2z^3 + 5abz^2 - abz + 2az + 2bz - 1)\cdot\Big(
	32a^4b^4z^8 + 32a^4b^3z^7 + 32a^3b^4z^7                                                                                                           \\
	        & - 8a^4b^3z^6 - 8a^3b^4z^6 + 400a^3b^3z^7 + 8a^4b^2z^6 - 68a^3b^3z^6 + 8a^2b^4z^6 - 4a^3b^3z^5                                           \\
	        & + 520a^3b^2z^6 + 520a^2b^3z^6 + a^3b^3z^4 - 160a^3b^2z^5 - 160a^2b^3z^5 + 15a^3b^2z^4 + 224a^3bz^5                                      \\
	        & + 15a^2b^3z^4 + 484a^2b^2z^5 + 224ab^3z^5 - 48a^3bz^4 - 215a^2b^2z^4 - 48ab^3z^4 + 32a^3z^4                                             \\
	        & + 32a^2b^2z^3 + 112a^2bz^4 + 112ab^2z^4 + 32b^3z^4  - 2a^2b^2z^2 - 24a^2bz^3 - 24ab^2z^3 + 2a^2bz^2 + 2ab^2z^2                          \\
	        & - 32abz^3 + 42abz^2 - 12abz - 16az^2 - 16bz^2 + ab + 8az + 8bz - a - b - 4z + 1 \Big)                                                   \\
	        & +                                                                                                                                       \\
\end{align*}
\begin{align*}
	 & \Big(
	4 a^{6} b^{6} z^{10}+4 a^{5} b^{5} \left(a +b +25\right) z^{9}
	-a^{4} b^{4} \left(a^{2} b +a \,b^{2}-a^{2}+21 a b -b^{2}-130 a -130 b -625\right) z^{8}                                                               \\
	 & + a^{3} b^{3} \left(a^{2} b^{2}-41 a^{2} b -41 a \,b^{2}+56 a^{2}-187 a b +56 b^{2}+1000 a +1000 b \right) z^{7}                                    \\
	 & + 4 a^{2} b^{2} \left(a^{3} b^{2}+a^{2} b^{3}-3 a^{3} b +4 a^{2} b^{2}-3 a \,b^{3}+2 a^{3}-109 a^{2} b -109 a \,b^{2}\right.                        \\
	 & \left.+2 b^{3}+150 a^{2}+250 a b +150 b^{2}\right) z^{6} -a b \left(a^{3} b^{3}-77 a^{3} b^{2}-77 a^{2} b^{3}+216 a^{3} b +397 a^{2} b^{2}\right.   \\
	 & \left.+216 a \,b^{3}-160 a^{3}-220 a^{2} b -220 a \,b^{2}-160 b^{3}\right) z^{5}                                                                    \\
	 & +\left(-5 a^{4} b^{3}-5 a^{3} b^{4}+21 a^{4} b^{2}+48 a^{3} b^{3}+21 a^{2} b^{4}-32 a^{4} b\right.                                                  \\
	 & \left. + 17 a^{3} b^{2}+17 a^{2} b^{3}-32 a \,b^{4}+16 a^{4} -48 a^{3} b -162 a^{2} b^{2}-48 a \,b^{3}+16 b^{4}\right) z^{4}                        \\
	 & +\left(-a^{3} b^{3}-23 a^{3} b^{2}-23 a^{2} b^{3}+48 a^{3} b +139 a^{2} b^{2} +48 a \,b^{3}-16 a^{3}-72 a^{2} b -72 a \,b^{2}-16 b^{3}\right) z^{3} \\
	 & +\left(2 a^{3} b^{2}+2 a^{2} b^{3}-6 a^{3} b -24 a^{2} b^{2}-6 a \,b^{3}+4 a^{3}+22 a^{2} b +22 a \,b^{2}+4 b^{3}-4 a^{2}-4 b^{2}\right) z^{2}      \\
	 & +\left(a^{2} b^{2}-a^{2} b -a \,b^{2}-7 a b +4 a +4 b \right) z  +\left(b -1\right) \left(a -1\right)  \Big).
\end{align*}

\pagebreak

\bibliography{three_walks-asym}{}
\bibliographystyle{unsrt}

\end{document}